\documentclass[a4paper,12pt]{article} 
\usepackage{amssymb} 
\usepackage{graphicx}      
 
\def \bea {\begin{eqnarray}} 
\def \eea {\end{eqnarray}}

\begin{document} 
 
\author{\textbf{A. J. Bracken }$^{\triangleleft }\thanks{\texttt{%
ajb@maths.uq.edu.au}}$, \textbf{\ D. Ellinas }$^{\triangleright }\thanks{%
\texttt{ellinas@science.tuc.gr}}$\textbf{\ and I. Smyrnakis }$%
^{\triangleright }\thanks{\texttt{smyrnaki@tem.uoc.gr}}$ \\ 
$^{\triangleleft }$Centre for Mathematical Physics, \\ 
\ Department of \ Mathematics, \ \\ 
\ University of Queensland, Brisbane 4072 Australia \\ 
\\ 
$^{\rhd }$Department of Sciences, Division of Mathematics, \\ 
\ Technical University of Crete, GR 731 00 \\ 
\ Chania Crete Greece\\ 
} 
\title{\textbf{Free Dirac evolution as a quantum random walk}} 
\maketitle 
 
\begin{abstract} 
Any positive-energy state of a free Dirac particle that is initially 
highly-localized, evolves in time by spreading at speeds close to the speed 
of light. This general phenomenon is explained by the fact that the Dirac 
evolution can be approximated arbitrarily closely by a quantum random walk, 
where the roles of coin and walker systems are naturally attributed to the 
spin and position degrees of freedom of the particle. Initially entangled 
and spatially localized spin-position states evolve with asymptotic 
two-horned distributions of the position probability,  
familiar from earlier studies of 
quantum walks. For the Dirac particle, the two horns travel apart at close 
to the speed of light. 
\end{abstract} 
 
\newpage 
 
\section{Introduction} 
 
The concept of a quantum random walk (QRW) has been widely discussed and 
extended in various directions \cite{kempe} since its introduction \cite%
{aharonov,meyer,ambainis4} and developement \cite%
{kempe1,fahri2,ambainis3,konno2,travaglione,sanders,dur,nielsenchuang}. 
Much of the interest has derived from an expectation that such a 
mathematically attractive idea should have important applications in 
quantum information theory, analogous to known applications of 
classical random walks (CRWs) in classical information science. 
 
On the other hand, CRWs also have many applications outside classical 
information theory, in a wide variety of areas of science where mathematical 
modelling is involved, so it should not be surprising if QRWs find 
applications outside quantum information theory. Here we describe such an 
application to relativistic quantum mechanics; for an earlier  
application in this field, see \cite{konno4}. 
 
The evolution in time of the state of a free Dirac particle, starting from a 
highly localized, positive-energy state, is a quantum process that has only 
recently been described fully \cite{bracken2}. There had been a widespread 
misapprehension that no relativistic particle with nonzero rest-mass $m$ 
could be localized much within its Compton wavelength $\lambda _{C}=\hbar /mc 
$, where $c$ is the speed of light. However it has now been shown \cite%
{bracken1} that there is no such difficulty for the Dirac particle if 
localization is characterized in terms of the Dirac position operator $%
\mathbf{x}$, by making $\Delta _{x}=\sqrt{\langle \mathbf{x}^{2}\rangle 
-\langle \mathbf{x}\rangle ^{2}}$ small while keeping the energy positive,  
and not by unrealistic attempts to 
restrict the domain of the wavefunction to a bounded region in configuration 
space. Arbitrarily precise localization, with $\Delta _{x}\ll \lambda _{C}$, 
is possible in the case of the free Dirac particle with positive energy. 
When the particle is localized in such an initial state, it has an 
associated uncertainty in energy $\Delta _{E}\gg mc^{2}$, and the subsequent 
evolution produces a probability density that spreads 
outwards in all directions at close to speed $c$. The graph of the evolving 
density along any axis through the centre of initial localization (see Fig. 
1 in \cite{bracken2}) shows a striking resemblance to the two-horned density 
found for a typical 1-dimensional QRW \cite{ambainis4}. For the Dirac 
particle, the horns are close to distance $ct$ from the starting point. We 
shall show that this is not a coincidence, and that the evolution of \emph{%
any} positive-energy state of a free Dirac particle moving in 1-dimension 
can be modelled arbitrarily closely as a QRW of the type described in detail 
by Ambainis \emph{et al.} \cite{ambainis4}, Konno \cite{konno1,konno3} and 
others. 
 
In addition to providing a somewhat surprising application of a QRW to a 
real process, this result provides some new insights as to the nature of the 
quantum walk itself. Until now the various proposed realization schemes for 
QRWs were typically based on the idea that the coin and walker degrees of 
freedom of the walk should be associated with two distinct quantum systems. 
These two systems were to be combined by means of some form of dynamical 
coupling-decoupling scheme. The present application shows that 
alternatively, a single quantum mechanical object such as the free Dirac 
particle --- by its very nature as a relativistic system with translational 
and spin degrees of freedom --- can be identified in the course of its time 
evolution with a quantum random walk. This natural occurence of a QRW, 
instead of some engineered realization, suggests that the question of its 
ontological status is still an interesting and open one. 
 
The present application draws attention to two other important features of 
QRWs that have been emphasized by others \cite{fahri2,childs1}. The first is that a 
QRW is a unitary evolution; the associated randomness is of the kind 
associated with every unitary quantum evolution. In particular, a QRW 
is typically reversible in time, unlike a CRW. The second feature is that a 
QRW is typically a ballistic process, associated with spreading at a 
constant speed, unlike diffusive CRWs, where spreading is proportional 
to the square root of the time. These two features are essential for the 
application that we describe below to the free Dirac evolution, which is a 
time-reversible process characterized by spreading near the speed of light. 
 
In what follows, we relate the evolution of Dirac's equation to that of a 
QRW based on the canonical Heisenberg algebra extended by the Dirac 
matrices. Then we construct and discuss the limiting probability 
distribution describing the translational spreading of an initial state. 
The paper concludes with 
some speculations about the physical reality of the quantum walk of the 
Dirac particle, and the possibility of detecting it experimentally. 
 
\section{QRW and free Dirac evolution} 
 
The free Dirac Hamiltonian operator for a particle with zero momentum along 
the $y$ and $z$ directions is 
 
\begin{equation} 
H({\hat{p}})=c\alpha \hat{p}+mc^{2}\beta \,,\quad {\hat{p}}=-i\hbar \,d/dx\,, 
\label{dirac_equation} 
\end{equation}%
acting on 4-component spinor wavefunctions $\Psi (x)$. Here we adopt a 
representation of the Dirac matrices with 
 
\begin{equation} 
\alpha =\sigma _{3}\otimes \sigma _{3}\,,\quad \beta =\sigma _{2}\otimes 
\mathbf{1}_{2}\,,  \label{dirac_matrices} 
\end{equation}%
where $\sigma _{i}$, $i=1,\,2,\,3$ are the usual Pauli matrices, and $%
\mathbf{1}_{2}$ is the $2\times 2$ unit matrix. In this representation the 
helicity (spin) operator associated with rotations about the $x$-axis is $%
\Sigma =\mathbf{1}_{2}\otimes \sigma _{3}$. From this point onwards we adopt 
the natural units $\hbar =c=m=1$. 
Recalling that only those solutions of Dirac's equation with positive energy 
describe physical electron states, we introduce the orthonormal 
positive-energy spinors in momentum space 
 
\begin{equation} 
u_{\pm }(p)=\frac{1}{2\sqrt{E(p)(E(p)+1)}}\left( 
\begin{array}{l} 
1+E(p)\pm p \\ 
\\ 
i(1+E(p)\mp p)%
\end{array}%
\right) \otimes e_{\pm },  \label{energy_spinors} 
\end{equation}%
where $e_{+}=(1,0)^{T}$, $e_{-}=(0,1)^{T}$ and $E(p)=\sqrt{p^{2}+1}$. 
These spinors satisfy the 
relations 
 
\begin{eqnarray} 
u_{\pm }(p)^{\dag }u_{\pm }(p)=1\,,\quad u_{\pm }(p)^{\dag }u_{\mp 
}(p)=0\,,\quad  \nonumber \\ 
\nonumber \\ 
H(p)u_{\pm }(p)=E(p)u_{\pm }(p)\,,\quad \Sigma u_{\pm}(p)=\pm \frac{1}{2} 
u_{\pm}(p)\,.  \label{spinor_normalization} 
\end{eqnarray}%
Now we can write an arbitrary positive-energy wavefunction (with zero $y$ 
and $z$ components of momentum) $\Psi _{0}(x)$ 
in terms of two arbitrary functions $f_{\pm }(p)$ as 
 
\begin{equation} 
\Psi _{0}(x)=\frac{1}{\sqrt{2\pi }}\int_{-\infty }^{\infty }e^{ipx}\left\{ 
f_{+}(p)u_{+}(p)+f_{-}(p)u_{-}(p)\right\} dp\,.  \label{psi_initial} 
\end{equation}%
Then 
\begin{equation} 
\int_{-\infty }^{\infty }\Psi _{0}(x)^{\dag }\Psi 
_{0}(x)\,dx=1\longleftrightarrow \int_{-\infty }^{\infty }\left\{ 
|f_{+}(p)|^{2}+|f_{-}(p)|^{2}\right\} \,dp=1\,.  \label{normalization} 
\end{equation} 
 
Suppose now that we choose a normalized positive-energy state with a 
definite helicity $+1/2$ and finite mean energy. Then 
 
\begin{equation} 
f_{+}(p)=f(p)\,,\quad f_{-}(p)=0\,,\quad \int_{-\infty }^{\infty 
}|f(p)|^{2}\,dp=1,  \label{f_choice} 
\end{equation}%
and 
 
\begin{equation} 
\langle \,H({\hat{p}}\,)\,\rangle =\int_{-\infty }^{\infty 
}E(p)|f(p)|^{2}\,dp=E_{0}<\infty \,.  \label{mean_energy} 
\end{equation} 
 
With $f_{-}(p)=0$, the action of $H(\,{\hat{p}}\,)$ in the second factor of 
the tensor product space in (\ref{dirac_matrices}) and (\ref{energy_spinors}%
) becomes trivial, as the second spinor remains constant at the value $e_{+}$%
. Thus the second factor space can be ignored, and we can consider $H(\,{%
\hat{p}}\,)$ to have the \emph{effective} form 
 
\begin{equation} 
H(\,{\hat{p}}\,)=\sigma _{3}{\hat{p}}+\sigma _{2} 
\end{equation}%
acting in the first factor space. In this first space, we write 
\begin{equation} 
(1\quad 0)^{T}=|+\rangle \,\quad (0\quad 1)^{T}=|-\rangle \,, 
\end{equation}%
so that the positive-energy spinor $u_{+}(p)$ in (\ref{energy_spinors}) 
takes the (effective) form 
 
\begin{equation} 
u_{+}(p)=\frac{1}{2\sqrt{E(p)(E(p)+1)}}\left\{ (1+E(p)+p)|+\rangle 
+i(1+E(p)-p)|-\rangle \right\} \,.  \label{effective_uplus} 
\end{equation} 
 
Next we consider a fixed, small time interval $\Delta t\ll 1/E_{0}\,.$ The 
(effective) unitary evolution operator for the Dirac particle can then be 
approximated over the time interval $\Delta t$ using the relations 
 
\begin{eqnarray} 
e^{-iH({\hat{p}})\Delta t} &=&V\,U+O([\Delta t]^{2})\,,  \nonumber \\ 
\nonumber \\ 
V &=&e^{-i\Delta t\sigma _{3} {\hat{p}}}\,,\quad U=e^{-i\Delta t\sigma _{2} 
} \,.  \label{approx_evolution1} 
\end{eqnarray}%
Here we see the appearance of the evolution operator $V\,U$ for a 
1-dimensional QRW \cite{ambainis4}, with $V$ enacting a step of length $%
\Delta t$ to the left or right along the $x$-axis (the ``walker space"), 
depending on the sign of $\sigma _{3}$, and with the reshuffling matrix $U$ 
representing the ``quantum coin toss" after each time interval of duration $%
\Delta t$. For a longer time $t=n\Delta t$, we have from (\ref%
{approx_evolution1}) 
 
\begin{equation} 
e^{-iH({\hat{p}})t}=(V\,U)^{n}+\mathrm{O}(\Delta t)\,, 
\label{approx_evolution2} 
\end{equation}%
and we see that the evolution of the state of the Dirac particle over any 
finite time $t$ can be obtained arbitrarily accurately by replacing the 
exact evolution operator by $(V\,U)^{n}$ and letting $n\rightarrow \infty $ 
and $\Delta t\rightarrow 0$ with $n\Delta t=t$. In other words,

\begin{eqnarray} 
\lim_{{{n\to\infty}\atop {\Delta t\to 0}}\atop {n\Delta t=t}} \,\left( 
e^{-i\Delta t \sigma_3{\hat p}}\, 
e^{-i\Delta t\sigma_2}\right)^n  
=e^{-iH({\hat p}) t}\,, 
\label{evolution3} 
\end{eqnarray} 
and we can 
emulate the Dirac evolution by the evolution of a QRW. If we are interested 
in particular in the evolution of the probability density along the $x$-axis 
for the electron, we need only go at each time $t$ to the asymptotic form of 
the QRW probability distribution for the ``walker" \cite%
{ambainis4,konno1,konno3}. 
 
It is important to note at this point that whereas the exact Dirac evolution 
operator $e^{-iH({\hat{p}})t}$ obviously preserves the positive energy 
condition for physically meaningful states, the same is not true of the 
approximate, QRW evolution $(V\,U)^{n}$. However, (\ref{approx_evolution2}) 
and (\ref{evolution3}) 
show that in the asymptotic limit described, the positive energy condition 
is respected. 
 
We close this section with the following remark. Rewriting the evolution 
operators as $V=|+\rangle \langle +| \,e^{-i\Delta t{\hat{p}}}+|-\rangle 
\langle -|\, e^{i\Delta t{\hat{p}}}$, $U=e^{-i\Delta t\sigma _{2}}$, we 
identify the type of quantum walk involved here as a Canonical Algebra QRW 
(CA-QRW) in the classification of \cite{ellinas}. \ In contrast to the 
Euclidean QRW, which takes place on the integers and whose evolution 
operator is constructed from the generators of the Euclidean algebra \cite%
{ellinas,ellinas1,ellinas2,ellinas3}, in the present case the generators of 
the canonical Heisenberg algebra --- position and momentum operators --- are 
used in the construction of a discrete walk on the $x$-coordinate axis. The close 
algebraic relationship between these two walks facilitates the solution of 
the time evolution in the present case, provided (as is done in next 
section) that we carefully discretize the coordinate-space (generalized) 
eigenfunctions, which unlike their Euclidean QRW counterparts, are not 
orthogonal. 
 
\section{Asymptotic solutions and localization} 
 
Let $\mathcal{H}$ denote the Hilbert space spanned by all vectors $%
|\Phi\rangle \otimes |\pm\rangle$, corresponding in the coordinate 
representation to normalizable 2-component wavefunctions  
$\Phi(x)|\pm\rangle$. 
Introduce a dense subspace $\mathcal{S}<\mathcal{H}$ consisting of all 
finite linear combinations of suitably regular vectors $|\Phi\rangle \otimes 
|\pm\rangle \in \mathcal{H}$, say all those corresponding to $%
\Phi(x)=P(x)\,e^{-\alpha x^2}$, where $P(x)$ is an arbitrary polynomial and $%
\alpha$ is some fixed positive constant. Then denote by $\mathcal{S}^{*}$ 
the space dual to $\mathcal{S}$ and, with the usual abuse of notation, 
consider $\mathcal{H}$ as a subspace of $\mathcal{S}^{*}$, so that we obtain 
the Gelf'and Triple (or Rigged Hilbert Space \cite{bohm}) 
\begin{eqnarray} 
\mathcal{S}<\mathcal{H}<\mathcal{S}^{*}\,.  \label{rigged} 
\end{eqnarray} 
The space $\mathcal{S}^{*}$ contains in particular the vectors $%
|x^{\prime}\rangle\otimes |\pm\rangle$, where $|x^{\prime}\rangle$ is the 
generalized eigenvector of the Dirac $x$-coordinate operator ${\hat q}$, 
\begin{equation} 
{\hat q}|x^{\prime}\rangle = x^{\prime}|x^{\prime}\rangle\,,  \label{q_evec} 
\end{equation} 
corresponding in the coordinate representation to $\delta(x-x^{\prime})$. 
 
The introduction of the time interval $\Delta t$ as in (\ref%
{approx_evolution1}), in turn defines a length interval $\Delta t$ on the $x$%
-axis (recall that $c=1$ now), and a corresponding direct-integral 
decomposition 
\begin{eqnarray} 
\mathcal{S}^{*}=\oplus \int_{-\Delta t/2}^{\Delta t/2}\, \mathcal{V}%
_{x_0}\,dx_0\,,  \label{Sdecompose} 
\end{eqnarray} 
where $\mathcal{V}_{x_0}<\mathcal{S}^{*}$ is spanned by all vectors of the 
form $|x_0+m\Delta t\rangle\otimes |\pm\rangle$, with $x_0\in (-\Delta 
t/2,\,\Delta t/2]$ fixed, and $m\in {\mathbb Z}$. We note at once that each $\mathcal{%
V}_{x_0}$ is invariant under the action of the QRW evolution operator $V\,U$%
, \textit{i.e.} 
\begin{eqnarray} 
V\,U\,\mathcal{V}_{x_0}< \mathcal{V}_{x_0}\,,  \label{invariance1} 
\end{eqnarray} 
because 
\begin{eqnarray} 
V\,|x_0+m\Delta t\rangle \otimes |\pm\rangle= |x_0+(m\mp 1)\Delta t\rangle 
\otimes |\pm\rangle\,.  \label{Vaction} 
\end{eqnarray} 
 
In order to describe the QRW evolution more fully, we now write the initial 
state with wave function as in (\ref{psi_initial}) and (\ref{f_choice}), as 
an entangled state of the walker and coin subsystems, 
 
\begin{equation} 
|\Psi _{0}\rangle \!\rangle =\int_{-\infty }^{\infty }\left\{ 
c_{+}(x)|x\rangle \otimes |+\rangle +c_{-}(x)|x\rangle \otimes |-\rangle 
\right\} \,dx\,,  \label{initial_state} 
\end{equation}%
where 
\begin{eqnarray} 
c_{+}(x) &=&\frac{1}{\sqrt{2\pi }}\int_{-\infty }^{\infty }\frac{1+E(p)+p}{2%
\sqrt{E(p)(E(p)+1)}}f(p)e^{ipx}dp\,,  \nonumber \\ 
c_{-}(x) &=&\frac{i}{\sqrt{2\pi }}\int_{-\infty }^{\infty }\frac{1+E(p)-p}{2%
\sqrt{E(p)(E(p)+1)}}f(p)e^{ipx}dp\,.  \label{coefficients} 
\end{eqnarray}%
Normalization of $|\Psi _{0}\rangle \!\rangle $ is satisfied because 
 
\begin{equation} 
\int_{-\infty }^{\infty }\left\{ |c_{+}(x)|^{2}+|c_{-}(x)|^{2}\right\} 
\,dx=1, 
\end{equation}%
as a consequence of (\ref{spinor_normalization}) and (\ref{normalization}). 
At this point we emphasize again that although, as is well known \cite%
{wigner}, $|x\rangle \otimes |\pm\rangle$ is \emph{not} a positive-energy 
(generalized) state, $|\Psi _{0}\rangle $ \emph{is} a positive-energy state,  
as a consequence of the 
particular form of the coefficients in (\ref{coefficients}). 
 
The expansion in (\ref{initial_state}) can be rewritten as 
\begin{eqnarray} 
|\Psi _{0}\rangle \!\rangle &=&\sum_{m\in \,{\mathbb Z}{}}\int_{-\Delta t/2}^{\Delta 
t/2}\left\{ c_{+}(x_0+ m\Delta t)\,|x_0+ m\Delta t\rangle \otimes |+\rangle 
\right.  \label{initial_state2} \\ 
&&\left. +c_{-}(x_0+ m\Delta t)\,|x_0+m\Delta t\rangle \otimes |-\rangle 
\right\} \,dx_{0}\,,  \nonumber 
\end{eqnarray}%
which is to be compared with (\ref{Sdecompose}). In view of the invariance 
of each $\mathcal{V}_{x_0}$ under the action of the QRW evolution, we can 
restrict our attention to that action on each substate%
\begin{eqnarray} 
|\Phi _{x_{0}}\rangle \!\rangle &=&\sum_{m\in \,{\mathbb Z}{}}\left\{ c_{+}(x_0+ 
m\Delta t)\,|x_0+ m\Delta t\rangle \otimes |+\rangle \right. 
\label{substate} \\ 
&&\left. +c_{-}(x_0+ m\Delta t)\,|x_0+ m\Delta t\rangle \otimes |-\rangle 
\right\}\sqrt{\Delta t}  \nonumber 
\end{eqnarray}%
with $x_{0}$ fixed, even though these substates are not normalizable, and 
are not positive-energy states. The point is that the general form of any 
such substate is preserved under the action of the QRW evolution $V\,U$, 
with no change in the value of $x_{0}$. The inclusion of the multiplicative 
factor $\sqrt{\Delta t}$ in last equation is for later convenience with the 
normalization. 
 
Consider firstly the action of $V$ on a general substate $|\Phi _{0}\rangle 
\!\rangle \in \mathcal{V}_{x_{0}},$ say one with $x_{0}=0$ for definiteness. 
\ We have%
\begin{eqnarray} 
V|\Phi _{0}\rangle \!\rangle &=&\sum_{m\in {\mathbb Z}\,{}}\,|m\Delta t\rangle 
\otimes \left\{ c_{+}((m-1)\Delta t)|+\rangle +c_{-}((m+1)\Delta t)|-\rangle 
\right\} \sqrt{\Delta t}  \nonumber \\ 
&=&\sum_{m\in {\mathbb Z}{}}\sum_{\alpha =\pm }[c_{\alpha }(m\Delta t)\,|(m+\alpha 
)\Delta t\rangle \otimes |\alpha \rangle ]\sqrt{\Delta t}  \nonumber \\ 
&\equiv &\left( E_{+}\otimes P_{+}+E_{-}\otimes P_{-}\right) \,|\Phi 
_{0}\rangle \!\rangle \,,  \label{V_action} 
\end{eqnarray}%
where 
\begin{equation} 
E_{\pm }|m\Delta t\rangle =|(m\pm 1)\Delta t\rangle \,,\quad P_{\pm }|\pm 
\rangle =|\pm \rangle \,,\quad P_{\mp }|\pm \rangle =0\,. 
\label{shift_actions} 
\end{equation}%
The action of $U$ on $|\Phi _{0}\rangle \!\rangle $ is easily seen from (\ref%
{approx_evolution1}), which implies that%
\begin{eqnarray} 
U\,|+\rangle &=&\cos (\Delta t)\,|+\rangle +\sin (\Delta t)\,|-\rangle \,, 
\nonumber \\ 
U\,|-\rangle &=&\cos (\Delta t)\,|-\rangle -\sin (\Delta t)\,|+\rangle \,. 
\label{U_action} 
\end{eqnarray}%
Combining (\ref{V_action}) and (\ref{U_action}), we see that 
\begin{equation} 
V\,U\,|\Phi _{0}\rangle \!\rangle =(E_{+}\otimes P_{+}U+E_{-}\otimes 
P_{-}U)\,|\Phi _{0}\rangle \!\rangle \,.  \label{VU_action} 
\end{equation} 
 
If we had taken $f_{+}(p)=0$, $f_{-}(p)=f(p)$ in (\ref{f_choice}), we would 
have written instead%
\begin{equation} 
u_{-}(p)=\frac{1+E(p)-p}{2\sqrt{E(p)(E(p)+1)}}|+\rangle +i\frac{1+E(p)+p}{2%
\sqrt{E(p)(E(p)+1)}}|-\rangle \,,  \label{alternative_spinor} 
\end{equation} 
and we would have obtained 
\begin{equation} 
|\Psi _{0}\rangle \!\rangle =\int_{-\infty}^{\infty} \left\{ 
c_{+}(x)|x\rangle \otimes |+\rangle +c_{-}(x)|x\rangle \otimes |-\rangle 
\right\} \,dx, 
\end{equation}%
where now 
\begin{eqnarray*} 
c_{+}(x) &=&\frac{1}{\sqrt{2\pi }}\int \frac{1+E(p)-p}{2\sqrt{E(p)(E(p)+1)}}%
f(p)e^{ipx}dp\,, \\ 
c_{-}(x) &=&\frac{i}{\sqrt{2\pi }}\int \frac{1+E(p)+p}{2\sqrt{E(p)(E(p)+1)}}%
f(p)e^{ipx}dp\,. 
\end{eqnarray*}%
Then, decomposing $|\Psi _{0}\rangle \!\rangle $ into substates $|\Phi 
_{x_{0}}\rangle \!\rangle $ as before, we would have obtained on a state of 
this general form, say one with $x_{0}=0$, that 
\begin{equation} 
V\,U|\Phi _{0}\rangle \!\rangle =\left\{ E_{-}\otimes P_{+}U+E_{+}\otimes 
P_{-}U\right\} |\Phi _{0}\rangle \!\rangle \,.  \label{alternative_VU_action} 
\end{equation}%
We will treat here the first case, as the second one can be treated 
similarly.  
 
To proceed we choose $-\pi \leq \phi <\pi $, and set 
\begin{eqnarray} 
|\phi/ \Delta t\rangle =\frac{1}{2\pi }\sum_{m\in \,{\mathbb Z}}e^{-im\phi }|m\Delta 
t\rangle\,,  \label{discrete_state_def} 
\end{eqnarray} 
so that 
\begin{eqnarray} 
E_{\pm }\,| \phi /\Delta t\rangle =e^{\pm i\phi }|\phi / \Delta 
t\rangle\,,\qquad |m\Delta t\rangle =\int_{-\pi }^{\pi }e^{im\phi }|\phi 
/\Delta t\rangle \,d\phi\,.  \label{discrete_state2} 
\end{eqnarray} 
Considering the evolution operator $V\,U$ acting as in (\ref%
{alternative_VU_action}), but now with $E_{\pm }$ diagonalized, we have 
\begin{equation} 
V\,U(\phi )=\left( e^{i\phi }P_{+}+e^{-i\phi }P_{-}\right) U\,. 
\label{E_diagonal_VU} 
\end{equation} 
The eigenvalues of this $2\times 2$ matrix with parameter $\phi$ are 
\begin{equation} 
\lambda _{\pm }(\phi )=\cos \phi \cos \Delta t\pm i\sqrt{1-\cos ^{2}\phi 
\cos ^{2}\Delta t}\,. 
\end{equation}%
 
Suppose that the corresponding eigenvectors are%
\begin{eqnarray} 
|v_{+}(\phi )\rangle &=&f_{++}(\phi )|+\rangle +f_{+-}(\phi )|-\rangle \,, 
\nonumber \\ 
|v_{-}(\phi )\rangle &=&f_{-+}(\phi )|+\rangle +f_{--}(\phi )|-\rangle \,. 
\label{evolution_eigenvectors1} 
\end{eqnarray} 
Then the eigenvectors of $V\,U$ are of the form $|\phi /\Delta t\rangle 
\otimes |v_{\pm}(\phi )\rangle$,  
with eigenvalues $\lambda _{\pm }(\phi ).$ Expanding $|\Phi 
_{0}\rangle \!\rangle $ in terms of these eigenvectors of $V\,U$ we get 
\begin{equation} 
|\Phi _{0}\rangle \!\rangle =\int_{-\pi }^{\pi }\left\{ g_{+}(\phi )\,|\phi 
/\Delta t\rangle \otimes |v_{+}(\phi )\rangle +g_{-}(\phi )\,|\phi /\Delta 
t\rangle \otimes |v_{-}(\phi )\rangle \right\} \sqrt{\Delta t}\,\,d\phi \,, 
\end{equation}%
where%
\begin{equation} 
g_{\pm }(\phi )=\sum_{m\in Z}\left\{ c_{+}(m\Delta t)f_{\pm +}^{\ast }(\phi 
)+c_{-}(m\Delta t)f_{\pm -}^{\ast }(\phi )\right\} e^{im\phi }\,. 
\end{equation}%
Hence 
\begin{eqnarray} 
|\Phi _{n}\rangle \!\rangle &\equiv &(V\,U)^{n}|\Phi _{0}\rangle \rangle 
=\int_{-\pi }^{\pi }\left\{ g_{+}(\phi )\lambda _{+}(\phi )^{n}|\phi /\Delta 
t\rangle \otimes |v_{+}(\phi )\rangle \right.  \nonumber \\ 
&&\left. +g_{-}(\phi )\lambda _{-}(\phi )^{n}|\phi /\Delta t\rangle \otimes 
|v_{-}(\phi )\rangle \right\} \sqrt{\Delta t}\,\,d\phi \,. 
\end{eqnarray} 
 
If we now denote by $X_{n}$ the random variable defining the ``walker 
position" after $n$ evolution steps, then we obtain for the ``quantum 
statistical moment" 
\begin{eqnarray} 
\langle (X_{n})^{k}\rangle &\equiv &\langle \!\langle \Phi _{n}|{\hat{q}}%
^{k}\otimes \mathbf{1}|\Phi _{n}\rangle \!\rangle =Tr_{S+T}\left( |\Phi 
_{n}\rangle \!\rangle \langle \!\langle \Phi _{n}|{\hat{q}}^{k}\otimes 
\mathbf{1}\right)  \nonumber \\ 
&=&Tr_{T}\left( (Tr_{S}|\Phi _{n}\rangle \!\rangle \langle \!\langle \Phi 
_{n}|){\hat{q}}^{k}\right) =Tr_{T}\left( \rho _{T}^{(n)}{\hat{q}}^{k}\right) 
, 
\end{eqnarray}%
where the expectation value has been expressed in terms of traces over the 
translational degree of freedom $(T)$ of the Dirac particle ---the walker 
system in the parlance of QRW --- and its spin $(S)$ --- the coin system for 
the QRW. This has allowed us to cast the ``quantum statistical moment" in 
terms of the reduced density operator $\rho _{T}^{(n)}=(Tr_{S}|\Phi 
_{n}\rangle \!\rangle \langle \!\langle \Phi _{n}|)$ which, as it provides 
all the necessary statistical information about the position of the Dirac 
particle, could have also been the main object of our mathematical 
investigation, as happens in most studies of QRWs. 
 
We proceed to determine the statistical moment of the translation operator, 
which takes the form 
 
\begin{eqnarray} 
\langle (X_{n})^{k}\rangle &=&\int_{-\pi }^{\pi }\left[ g_{+}^{\ast }(\phi 
)\lambda _{+}^{\ast }(\phi )^{n}(-i\partial _{\phi })^{k}\left\{ g_{+}(\phi 
)\lambda _{+}(\phi )^{n}\right\} \right.  \nonumber \\ 
&&\left. +g_{-}^{\ast }(\phi )\lambda _{-}^{\ast }(\phi )^{n}(-i\partial 
_{\phi })^{k}\left\{ g_{-}(\phi )\lambda _{-}(\phi )^{n}\right\} \right] \,%
\frac{d\phi }{2\pi }\,(\Delta t)^{k+1}, 
\end{eqnarray}%
or equivalently%
\begin{eqnarray} 
\langle (X_{n})^{k}\rangle &=&\left( n\Delta t\right) ^{k}\int_{-\pi }^{\pi 
}\left\{ \left\vert g_{+}(\phi )\right\vert ^{2}\left( -i\lambda 
_{+}^{\prime }(\phi )/\lambda _{+}(\phi )\right) ^{k}\right.  \nonumber \\ 
&&\left. +\left\vert g_{-}(\phi )\right\vert ^{2}\left( -i\lambda 
_{-}^{\prime }(\phi )/\lambda _{-}(\phi )\right) ^{k}\right\} \,\frac{%
d\phi }{2\pi }\,\Delta t+{\rm O}(n\Delta t)^{k-1}. 
\end{eqnarray}%
Hence as $n\rightarrow \infty $, $\Delta t\rightarrow 0$, with $t=n\Delta t$ 
\ large, we have that 
\begin{eqnarray} 
\left\langle \left( X_{n}/n\Delta t\right) ^{k}\right\rangle &\sim 
&\int_{-\pi }^{\pi }\left\{ \left\vert g_{+}(\phi )\right\vert ^{2}\left( 
-i\lambda _{+}^{\prime }(\phi )/\lambda _{+}(\phi )\right) ^{k}\right. \\ 
&&\left. +\left\vert g_{-}(\phi )\right\vert ^{2}\left( -i\lambda 
_{-}^{\prime }(\phi )/\lambda _{-}(\phi )\right) ^{k}\right\} \,\frac{d\phi 
}{2\pi }\,\Delta t.  \nonumber 
\end{eqnarray} 
 
This admits the following interpretation \cite{grimmett1}: \ we can take as 
a random variable, a function $Y$ from $\Omega =S^{1}\times \{+,-\}$ to the 
reals, with $Y=-i\lambda _{+}^{\prime }(\Phi )/\lambda _{+}(\Phi )$ on $%
S^{1}\times \{+\}$, and $Y=-i\lambda _{-}^{\prime }(\Phi )/\lambda _{-}(\Phi 
)$ on $S^{1}\times \{-\}.$ \ Here $\Phi :\Omega \rightarrow R$ is a random 
variable which projects on the circle $S^{1}$ with measure $\left\vert 
g_{+}(\phi )\right\vert ^{2}\Delta t\,(d\phi /2\pi )$ on $S^{1}\times \{+\}$%
, and measure $\left\vert g_{-}(\phi )\right\vert ^{2}\Delta t(d\phi /2\pi )$ 
on $S^{1}\times \{-\}.$ \ Since in the above limit, all the moments of $%
X_{n}/n\Delta t$ agree with all the moments of $Y$, and the support of $%
X_{n}/n\Delta t$ is compact, it follows that $X_{n}/n\Delta t$ converges 
weakly to $Y$. \ Hence we have that 
 
\begin{eqnarray} 
\lim_{n\Delta t = t\rightarrow \infty }P(y_{1}\leq X_{n}/n\Delta t\leq 
y_{2})=P(y_{1}\leq Y\leq y_{2})  \nonumber \\ \nonumber\\ 
=\int_{T_{+}}\left\vert g_{+}(\phi )\right\vert ^{2}\Delta t\,\frac{d\phi 
}{2\pi }\qquad +\int_{T_{-}}\left\vert g_{-}(\phi )\right\vert ^{2}\Delta t\,%
\frac{d\phi }{2\pi },  \label{limitpd} 
\end{eqnarray}%
where the intervals of integration are $T_{\pm }= y_{1}\leq (-i\lambda 
_{\pm }^{\prime }(\Phi )/\lambda _{\pm }(\Phi ))\leq y_{2}$. Then it follows 
that in order to determine the long time position distribution we need only 
determine $g_{\pm }(\phi )$ and $\lambda _{\pm }(\phi ).$ \ 
 
Suppose now that we specialize to the case of a highly localized initial 
electron state \cite{bracken2} with $f(p)=f_{\nu }(p)=e^{-p^{2}/2\nu ^{2}}/(%
\sqrt{\nu \sqrt{\pi }})$, where $\nu $ is large and positive and quantifies 
the extent of the localization of the Dirac particle's initial state --- the 
larger is $\nu$, the sharper is the initial localization. \ As $\nu $ 
approaches infinity one has that%
\begin{eqnarray} 
c_{+}(x) &\sim &\frac{1}{\sqrt{2\pi }}\sqrt{\frac{\nu }{\sqrt{\pi }}}%
\int_{0}^{\infty }e^{i\nu px}e^{-p^2/2\nu^2}dp, \\ 
c_{-}(x) &\sim &\frac{i}{\sqrt{2\pi }}\sqrt{\frac{\nu }{\sqrt{\pi }}}%
\int_{-\infty }^{0}e^{i\nu px}e^{-p^2/2\nu^2}dp. 
\end{eqnarray}%
\ Note that in the limit $\nu \rightarrow \infty $, $\int_{-\infty }^{\infty 
}|c_{+}(x)|^{2}dz=\int_{-\infty }^{\infty }|c_{-}(x)|^{2}dz=1/2$. \ If we 
now make $\nu \Delta t$ small by taking $\Delta t$ small enough, then 
\begin{eqnarray} 
g_{\pm }(\phi ) &\sim &i\sqrt{2\sqrt{\pi }/(\nu \Delta t^{2})}\,e^{-\phi 
^{2}/(2\nu ^{2}\Delta t^{2})}f_{\pm +}^{\ast }(\phi )\qquad if\quad \phi >0 
\nonumber \\ 
&\sim &\sqrt{2\sqrt{\pi }/(\nu \Delta t^{2})}\,e^{-\phi ^{2}/(2\nu 
^{2}\Delta t^{2})}f_{\pm -}^{\ast }(\phi )\qquad if\quad \phi <0. 
\label{g_values} 
\end{eqnarray}%
Also 
\begin{equation} 
-i\lambda _{\pm }^{\prime }(\phi )/\lambda _{\pm }(\phi )=\pm \frac{\sin 
\phi \cos \Delta t}{\sqrt{1-\cos ^{2}\Delta t\cos ^{2}\phi }}\equiv \pm 
h(\phi ).  \label{h_values} 
\end{equation}%
Note also that $|g_{+}(\phi )|^{2}+|g_{-}(\phi )|^{2}=2\sqrt{\pi }e^{-\phi 
^{2}/\nu ^{2}\Delta t^{2}}/(\nu \Delta t^{2}).$ To compute the asymptotic 
distribution we need to compute the integrals 
\begin{eqnarray} 
I_{1} &=&\sum_{i}\int_{h_{i}^{-1}[y_{1},y_{2}]}\left\vert g_{+}(\phi 
)\right\vert ^{2}\Delta t\frac{d\phi }{2\pi }  \nonumber \\ 
&=&\frac{1}{2\pi }\sum_{i}\int_{y_{1}}^{y_{2}}\frac{1}{\left\vert 
h_{+}^{\prime }(h_{i}^{-1}(y))\right\vert }\left\vert 
g_{+}(h_{i}^{-1}(y))\right\vert ^{2}\Delta t\,dy  \label{int1} 
\end{eqnarray}%
and 
 
\begin{eqnarray} 
I_{2} &=&\sum_{i}\int_{h_{i}^{-1}[-y_{2},-y_{1}]}\left\vert g_{-}(\phi 
)\right\vert ^{2}\Delta t\frac{d\phi }{2\pi }  \nonumber \\ 
&=&\frac{1}{2\pi }\sum_{i}\int_{y_{1}}^{y_{2}}\frac{1}{\left\vert 
h_{-}^{\prime }(h_{i}^{-1}(-y))\right\vert }\left\vert 
g_{-}(h_{i}^{-1}(-y))\right\vert ^{2}\Delta t\,dy,  \label{int222} 
\end{eqnarray}%
where the index $i$ labels the local inverses \ of the function $h.$ \ 
Because of (\ref{g_values}), the only inverse relevant to leading order is 
the one that keeps $\phi $ close to zero. For this inverse, $%
h_{i}^{-1}(-y)=-h_{i}^{-1}(y)$. Furthermore, and realizing that $|v_{\pm 
}(\phi )\rangle ^{\ast }=|v_{\mp }(-\phi )\rangle $, we can show that $%
|g_{\pm }(-\phi )|^{2}=|g_{\pm }(\phi )|^{2}$. A direct computation then 
gives 
\begin{equation} 
I_{1}+I_{2}=\frac{\Delta t\sin \Delta t}{2\pi }\int_{y_{1}}^{y_{2}}\frac{1}{%
(1-y^{2})\sqrt{\cos ^{2}\Delta t-y^{2}}}\frac{2\sqrt{\pi }}{\nu \Delta t^{2}}%
e^{-\left( h_{i}^{-1}(y)\right) ^{2}/\nu ^{2}\Delta t^{2}}\,dy. 
\label{sum_evaluation} 
\end{equation}%
If we set $y=$ $h(\phi ),$ then for all inverses $\phi _{i}$ we have $\cos 
^{2}\phi _{i}$ =$(\cos ^{2}\Delta t-y^{2})/\cos ^{2}\Delta t(1-y^{2})$. 
Since for our inverse the value of $\phi _{i}$ is small we have 
 
\begin{eqnarray} 
\left( h_{i}^{-1}(y)\right) ^{2}=\phi _{i}^{2}\simeq \sin ^{2}\phi 
_{i}&=&1-(\cos ^{2}\Delta t-y^{2})/{(\cos ^{2}\Delta t)(1-y^{2})}  \nonumber 
\\ 
&=&y^{2}\sin ^{2}\Delta t/\cos ^{2}\Delta t(1-y^{2})\,.  \label{hi_formula} 
\end{eqnarray} 
Taking the limit $\Delta t\rightarrow 0$ we arrive at the asymptotic 
distribution function associated with the random variable $X_{n}/n\Delta 
t\sim Y$, 
\begin{eqnarray} 
P(y_{1}\leq Y\leq y_{2})=\lim_{\Delta t\rightarrow 0}\left( 
I_{1}+I_{2}\right)  
=\int_{y_1}^{y_2}F(y)\,d,, 
\nonumber\\ 
\nonumber\\ 
F(y)=\frac{1}{\nu \sqrt{\pi }}\,\frac{1}{%
(1-y^{2})^{3/2}}e^{-y^{2}/\nu ^{2}(1-y^{2})}\,. 
\end{eqnarray} 
In Fig. 1 we plot this two-horned 
probability distribution for three values of the localization parameter $\nu$, 
and recognize it as the 1-dimensional 
analogue of the result obtained for the Dirac particle in three dimensions in \cite{bracken2}, 
Eqn (3.1). 
 
\begin{figure} 
\centering     
\includegraphics[angle=270,scale=0.6]{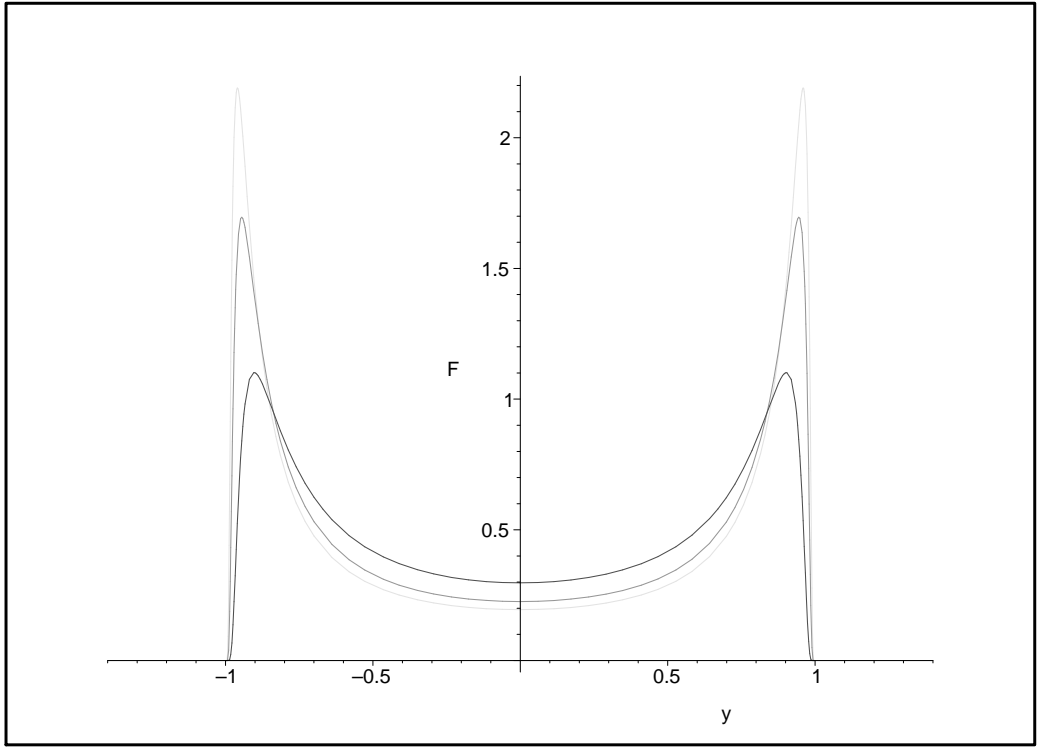}     
\caption{The asymptotic position probability density function, with 
localization parameter $\nu =1.9$, $2.5$ and $2.9$. As $\nu$ increases, the plots become more 
sharply peaked near the ends of the interval.}   
\end{figure}

\section{Discussion} 
 
It has been shown that the 1-dimensional Dirac evolution of a state with 
positive energy and definite spin is equivalent to a QRW, in the 
limit of small positional steps and a large number of iterations. An initial 
state that is highly localized, with all but one momentum component set to 
zero, spreads in the remaining direction at a speed that almost surely 
approaches the speed of light as the initial localization increases. 
 
This relationship between the Dirac evolution and a QRW leads to 
the intriguing speculation that at some small space-time scale, there may 
really be a QRW defining the evolution of states of the relativistic electron,  
and that it is the Dirac 
evolution that is only a large scale approximation. One way to test this 
would be to make very precise measurements of the spreading characteristics 
of initially highly localized electron states over short distances. 
Comparison with the characteristics that are typical for a QRW, in 
particular the shape of the position probability distribution at early times,  
may reveal whether or not there 
is indeed a QRW underlying an approximate Dirac evolution. 
 
It is tempting to speculate further that there may be some deep relationship 
between such an underlying QRW and the \emph{Zitterbewegung} of  
the relativistic electron, as first discussed by Schr\"odinger  
\cite{schrodinger}. This awaits further study. 
 
\bigskip \noindent \textit{Acknowledgments:} A.J.B acknowledges the support 
of Australian Research Council Grant DP0450778. The work of D. E. and I. S. 
was supported by the EPEAEK research programme ``Pythagoras II."

\bigskip 
 
\bigskip 
 
\bigskip Figure caption: 
 
Figure 1. The asymptotic position probability density function, with 
localization parameter $\nu =1.9$, $2.5$ and $2.9$. As $\nu$ increases, the plots become more 
sharply peaked near the ends of the interval.   
 
\bigskip 
 
\end{document}